\documentclass[aps,preprint,prd,nofootinbib]{revtex4-1}

\usepackage{graphicx}
\usepackage{epsfig}
\usepackage{epstopdf}
\usepackage{amsmath}
\usepackage{amsfonts}
\usepackage{amssymb}

\usepackage{booktabs}
\usepackage{txfonts}
\usepackage{color}
\allowdisplaybreaks

\begin{document}

\title{Electromagnetic form factors of spin $1/2$ doubly charmed baryons}

\author{Astrid N.~Hiller Blin}\email{hillerbl@uni-mainz.de}
\affiliation{Institut f\"ur Kernphysik \& PRISMA Cluster of Excellence, Johannes Gutenberg Universit\"at, D-55099 Mainz, Germany}

\author{Zhi-Feng Sun}\email{sunzhif09@lzu.edu.cn}
\affiliation{Research Center for Hadron and CSR Physics, Lanzhou University $\&$ Institute of Modern Physics of CAS, Lanzhou 730000, China}
\affiliation{School of Physical Science and Technology, Lanzhou University,
Lanzhou 730000, China}

\author{M. J. Vicente Vacas}\email{vicente@ific.uv.es}
\affiliation{Departamento de F\'{\i}sica Teorica and Instituto de Fisica Corpuscular (IFIC),
Centro Mixto UVEG-CSIC, Valencia E-46071, Spain}

\date{\today}

\begin{abstract}

We study the electromagnetic form factors of the doubly charmed baryons, using covariant chiral perturbation theory within the extended on-mass-shell (EOMS) scheme. Vector-meson contributions are also taken into account. We present results for the baryon magnetic moments, charge and magnetic radii.
While some of the chiral Lagrangian parameters could be set to values determined in previous works, the available lattice results for $\Xi_{cc}^+$ and $\Omega_{cc}^+$ only allow for robust constraints on the low-energy constant (LEC) combination, $c_{89}(=-\frac{1}{3}c_8+4c_9)$.
The couplings of the doubly charmed baryons to the vector mesons have been  estimated assuming the Okubo--Zweig--Iizuka (OZI) rule.
We also give the expressions for the form factors of the double beauty baryons  considering the masses predicted in the framework of quark models.
A comparison of our results with those obtained in heavy baryon chiral perturbation theory (HBChPT) at the same chiral order is made.

\end{abstract}

\maketitle

\section{Introduction}

The recent announcement of the observation of the $\Xi_{cc}^{++}$ particle by the LHCb
collaboration~\cite{Aaij:2017ueg}  has revived the interest on the physics of doubly heavy baryons. Up to now, the experimental evidence for baryons with two heavy quarks was marginal. Only one $cc$ baryon, $\Xi_{cc}^+(3520)$, had been included in the  Review of Particle Physics by the PDG~\cite{Patrignani:2016xqp} and was labeled with one star.
The $\Xi_{cc}^+(3520)$ baryon was first observed by the SELEX collaboration in
the $\Lambda_c^+K^-\pi^+$ channel~\cite{Mattson:2002vu} and later corroborated in the $pD^+K^-$ one~\cite{Ocherashvili:2004hi}. However, neither BABAR~\cite{Aubert:2006qw}, nor BELLE~\cite{Kato:2013ynr}, nor LHCb~\cite{Aaij:2013voa} could confirm the existence of this state.

On the other hand, the mass of $\Xi_{cc}^{++}$ measured by LHCb  is greater by more than 100 MeV than that of the $\Xi_{cc}^+(3520)$ particle. This large splitting, if confirmed, would suggest that the two states are not isospin partners~\cite{Karliner:2017gml} and thus could belong to different multiplets.

The scarce and conflicting experimental information has not deterred the theoretical research on the topic. Since the early works predicting the existence of doubly charmed baryons~\cite{Gaillard:1974mw,DeRujula:1975qlm}, soon after the discovery of hidden ($c\bar c$) charm states, these particles have been studied in quark models, QCD sum rules, lattice simulations, effective theories implementing heavy quark spin symmetry, etc. See, e.g., Refs.~\cite{Richard:2002nn,Cheng:2015rra}.
Most of those works address the spectroscopy of these baryons. Besides,
 other matters such as their decays~\cite{Kiselev:1998sy,Guberina:1999mx,Chang:2007xa,Karliner:2014gca} and electromagnetic properties have  been abundantly studied. For instance, diverse quark
models~\cite{Lichtenberg:1976fi,Jena:1986xs,
SilvestreBrac:1996bg, JuliaDiaz:2004vh, Faessler:2006ft,Albertus:2006ya,Sharma:2010vv,Shah:2017liu}, the MIT bag model
\cite{Bose:1980vy, Bernotas:2012nz}, the skyrmion model \cite{Oh:1991ws},
%heavy baryon chiral perturbation theory
HBChPT~\cite{Li:2017cfz} and ChPT within the EOMS scheme~\cite{Liu:2018Poster} have been applied to study the magnetic moments of doubly charmed baryons. Both, the magnetic and the electric form factors (FF) have also been investigated in lattice QCD~\cite{Can:2013zpa, Can:2013tna,Can:2015exa}. As expected, the lattice results show that the doubly charmed baryons have smaller radii  than the singly charmed ones and than those composed of only light quarks such as the proton. However, there is some tension between the lattice results for magnetic moments  and those determined in other theoretical models~\cite{Can:2013tna}.
 Here, we focus on these FF, which are a quite interesting probe, as they offer an insight on the hadron structure and how its constituents are distributed.

Our work is based on chiral perturbation theory (ChPT)~\cite{Scherer:2002tk,Gasser:1983yg, Gasser:1984gg,Gasser:1987rb}, which provides a model independent and systematic framework to study the non-perturbative regime of the strong interaction at low energies or for soft probes. Furthermore, it is well suited to analyse the lattice data at quark(meson) masses different from the physical ones.
 The ChPT results are systematically arranged as an expansion in powers of the Goldstone boson masses and the external (small) momenta. The corresponding power counting involves some difficulties when baryon loops are included in the calculation and different methods have been developed to overcome this issue such as HBChPT~\cite{Jenkins:1990jv}, heavy hadron (HH)ChPT, using similar techniques~\cite{Cho:1992gg,Cho:1992nt}, and the covariant approaches: infrared~\cite{Becher:1999he} and EOMS~\cite{Fuchs:2003qc}. All these schemes have been widely and successfully used to investigate
the electromagnetic structure of light
baryons~\cite{Caldi:1974ta,Gasser:1987rb,Krause:1990xc,
Jenkins:1992pi,Meissner:1997hn,Bos:1997rw,Durand:1997ya,Donoghue:1998bs,Puglia:1999th,Mojzis:1999qw,Puglia:2000jy,Kubis:2000zd,
Kubis:2000aa,Holstein:2000yf,Savage:2001dy,Chen:2001yi,Borasoy:2002jv,Arndt:2003ww,Beane:2004tw,Detmold:2004ap,
Bernard:2007zu,Lacour:2007wm,Geng:2008mf,Geng:2009hh,FloresMendieta:2009rq,Jiang:2009jn,Ahuatzin:2010ef,Wang:2014yra,
Xiao:2018rvd,Li:2016ezv,Blin:2017hez}. There are also some
calculations of the electromagnetic properties of baryons with a
single heavy quark in
HHChPT~\cite{Cho:1992nt,Savage:1994zw,Banuls:1999mu,Tiburzi:2004mv,Wang:2018gpl}.
Recently, the magnetic moments of doubly heavy baryons with spin
$\frac{1}{2}$ and $\frac{3}{2}$ have been studied in HBChPT
\cite{Li:2017cfz,Meng:2017dni}. Here, we use the
covariant EOMS framework instead and we also calculate the
electric and magnetic radii of the spin $\frac{1}{2}$ triplet.
The manifestly Lorentz invariant EOMS scheme has been found to deliver
a better chiral convergence than the other schemes for most
observables~\cite{Geng:2013xn} and in particular for the magnetic
moment of the light baryons~\cite{Geng:2008mf,Geng:2009ys}. Although
the HB techniques are expected to work better the larger the
baryon mass is, the differences with the covariant calculation are
not negligible. This point is also explored both for
di-charm and di-bottom baryons.

Our work is organized as follows. In Section II, the effective Lagrangian describing the interaction of doubly-heavy baryons and Goldstone bosons is given. The form factors of doubly heavy baryons are introduced in Section III and the results are shown in Section IV. Finally, summary and conclusions are given in Section V.

\section{The effective Lagrangian}

\subsection{Interaction with light pseudoscalar mesons}
The effective Lagrangian describing the interaction of double-charm baryons and the Goldstone bosons up to second order was constructed in Refs.~\cite{Sun:2014aya,Sun:2016wzh}. It can be written as
\footnote{In Refs. \cite{Sun:2014aya,Sun:2016wzh}, the $c_8$ terms
involves $f_{\mu\nu}^+$ instead of $\hat{f}_{\mu\nu}^+$. However, both formulations are equivalent, and the Lagrangian in Eq. \eqref{eq:L2} can be obtained by a redefinition of $c_9$ in Refs.~\cite{Sun:2014aya,Sun:2016wzh}. In addition, note that the $c_6$ term  in Eq. (14) of Ref.
\cite{Sun:2014aya} is hermitian.}
\begin{eqnarray}
\mathcal{L}^{(1)}&=&\bar{\psi}(i{D\!\!\!\slash}-m+\frac{g_A}{2}\gamma^\mu\gamma_5u_\mu)\psi,\label{eq:L1}\\
\nonumber
\mathcal{L}^{(2)}&=&c_1\bar{\psi}\langle
\chi_+\rangle\psi-\left\{\frac{c_2}{8m^2} \bar{\psi}\langle u_\mu
u_\nu \rangle \{ D^\mu,D^\nu\}\psi+h.c.\right\}\\\nonumber
&&-\left\{\frac{c_3}{8m^2} \bar{\psi}\{ u_\mu, u_\nu\}  \{
D^\mu,D^\nu\}\psi+h.c.\right\}+\frac{c_4}{2}\bar{\psi}\langle
u^2\rangle\psi\\\nonumber &&+\frac{c_5}{2}\bar{\psi} u^2\psi+
\frac{ic_6}{4}\bar{\psi}\sigma^{\mu\nu}[u_\mu,u_\nu]\psi+c_7\bar{\psi}\hat{\chi_+}\psi\\
&&+\frac{c_8}{8m}\bar{\psi}\sigma^{\mu\nu}\hat{f}^+_{\mu\nu}\psi+\frac{c_9}{8m}\bar{\psi}\sigma^{\mu\nu}\langle
f^+_{\mu\nu}\rangle\psi.\label{eq:L2}
\end{eqnarray}
The relevant pieces of the Lagrangian of order three can be obtained
by considering chiral, parity and charge conjugation symmetry. There
are two terms contributing to the electromagnetic form factors,
\begin{eqnarray}
\mathcal{L}^{(3)}&=&\left\{\frac{i}{2m}d_1\bar{\psi}[D^\mu,\hat{f}^+_{\mu\nu}]D^\nu\psi+h.c.\right\}+\left\{\frac{2i}{m}d_2\bar{\psi}[D^\mu,\langle
f^+_{\mu\nu} \rangle]D^\nu\psi+h.c.\right\}+...\,. \label{eq:L3}
\end{eqnarray}
The Lagrangians for the double beauty baryons are analogous, only modifying $m$, their mass in the chiral limit, and the coupling constants.
In these equations, $U=u^2$, which incorporates the pseudoscalar meson field, is defined as
\begin{eqnarray}
U=u^2=\exp\left(i\frac{\phi(x)}{F}\right),
\end{eqnarray}
where $\phi(x)$ is expressed as
\begin{eqnarray}
\phi(x)=\left(
          \begin{array}{ccc}
            \pi^0+\frac{1}{\sqrt{3}}\eta & \sqrt{2}\pi^+ & \sqrt{2}K^+ \\
            \sqrt{2}\pi^- & -\pi^0+\frac{1}{\sqrt{3}}\eta & \sqrt{2}K^0 \\
            \sqrt{2}K^- & \sqrt{2}\bar{K^0} & -\frac{2}{\sqrt{3}}\eta \\
          \end{array}
        \right).
\end{eqnarray}
The doubly-heavy baryon field $\psi$ with spin $\frac{1}{2}$ is a
column vector in the flavor space, i.e.
%\begin{eqnarray}
%\psi &=&\left(
%          \begin{array}{c}
%            \Xi_{\mathcal{QQ}}^{++} \\
%            \Xi_{\mathcal{QQ}}^{+} \\
%            \Omega_{\mathcal{QQ}}^+ \\
%          \end{array}
%        \right),
%\end{eqnarray}
%%% charges were not correct for beauty!
\begin{eqnarray}
\psi &=&\left(
          \begin{array}{c}
            \Xi_{\mathcal{QQ}}^u \\
            \Xi_{\mathcal{QQ}}^d \\
            \Omega_{\mathcal{QQ}}^s \\
          \end{array}
        \right),
\end{eqnarray}
where the subscript $\mathcal{Q}$ denotes the charm or beauty quark. In
Eqs. \eqref{eq:L1} and \eqref{eq:L2}, $\chi$, $\chi_\pm$, $f_{\mu\nu}$,
$f_{\mu\nu}^\pm$, $u_\mu$, $\Gamma_\mu$, $D_\mu$ have the following
definitions
\begin{eqnarray}
\chi&=&diag(M_\pi^2,M_\pi^2,2M_K^2-M_\pi^2),\\
\chi_{\pm}&=&u^\dag\chi u^\dag\pm u\chi^\dag u,\\
f_{\mu\nu}&=&-eQ\partial_\mu A_\nu+eQ\partial_\nu A_\mu,\\
f^+_{\mu\nu}&=&u^\dag f_{\mu\nu} u+ uf_{\mu\nu} u^\dag,\\
u_\mu&=&i[u^\dag(\partial_\mu+eQiA_\mu)u-u(\partial_\mu+eQiA_\mu)u^\dag],\\
\Gamma_\mu&=&\frac{1}{2}[u^\dag(\partial_\mu+eQiA_\mu)u+u(\partial_\mu+eQiA_\mu)u^\dag],\\
D_\mu&=&\partial_\mu+\Gamma_\mu,
\end{eqnarray}
with $A_\mu$ the photon field. For the double-charm baryons
$Q=diag(2,1,1)$, while for the double-beauty baryons $Q=diag(0,-1,-1)$.
For a $3\times 3$ matrix $A$ in flavor space, we define
$\hat{A}=A-\frac{1}{3}\langle A\rangle$ with $\langle A\rangle$ the
trace of $A$.

The interaction Lagrangian describing the Goldstone-boson interaction with a photon can be extracted from the leading-order meson Lagrangian
\begin{eqnarray}
\mathcal{L}&=&\frac{F^2}{4}\textup{Tr}[\tilde{D}_\mu U(\tilde{D}^\mu U)^\dag]\label{eq:14}
\end{eqnarray}
as follows
\begin{eqnarray}
\mathcal{L}_{\phi\phi\gamma}&=&\frac{ie}{2}\textup{Tr}[(\phi
\partial_\mu \phi-\partial_\mu \phi \phi)Q_l]A^\mu.\label{eq:LM}
\end{eqnarray}
In Eq. \eqref{eq:14}, $\tilde{D}_\mu U=\partial_\mu U+ieQ_lA_\mu
U-ieUQ_lA_\mu$ with $Q_l=diag(2/3,-1/3,-1/3)$.

\subsection{Interaction with vector mesons}

\begin{figure}
\centering
\includegraphics[width=0.20\linewidth]{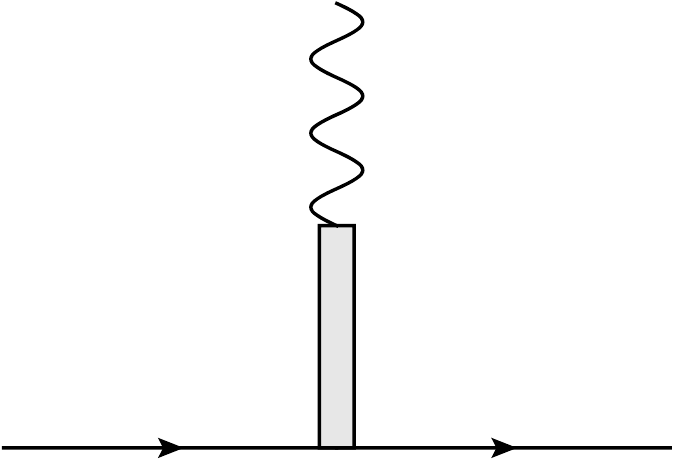}
\caption{Feynman diagram of the vector-meson contribution to the form factor.
\label{fig:vector}}
\end{figure}

It is well known, in the case of light baryons, that the consideration of a pseudoscalar meson cloud plus contact terms, even up to order $O(q^4)$, is not sufficient to provide a precise description of the electromagnetic form factors in ChPT~\cite{Kubis:2000zd,Kubis:2000aa,Blin:2017hez}. This is especially true for the $Q^2$ dependence and thus the charge and magnetic radii. The reason is the importance of the contribution of vector meson mechanisms,
see Fig.~\ref{fig:vector}. We expect a similar situation for the case of heavy baryons.

Therefore, in order to model the behavior of the form factors at moderate
momentum transfers, the vector-meson contributions are also included.
In the case of ideal mixing of the vector-meson singlet and octet, the Lagrangian of the coupling of doubly-heavy baryons to the vector mesons has the following structure
\begin{eqnarray}
\mathcal{L}_{VBB}&=&(\bar{\Xi}_{\mathcal{QQ}}^{++},\bar{\Xi}_{\mathcal{QQ}}^{+})\left(g_{v}^{\Xi_\mathcal{QQ}}\gamma^\mu+g_{t}^{\Xi_\mathcal{QQ}}
\frac{\sigma^{\mu\nu}\partial_\nu}{2m_B}\right)\left(
                                               \begin{array}{cc}
                                                 \frac{1}{\sqrt{2}}\rho^0+\frac{1}{\sqrt{2}}\omega & \rho^+ \\
                                                 \rho^- & -\frac{1}{\sqrt{2}}\rho^0+\frac{1}{\sqrt{2}}\omega \\
                                               \end{array}
                                             \right)
_\mu\left(
                                                          \begin{array}{c}
                                                            \Xi_{\mathcal{QQ}}^{++} \\
                                                            \Xi_{\mathcal{QQ}}^{+} \\
                                                          \end{array}
                                                        \right)
\nonumber\\
&&+\bar{\Omega}_{\mathcal{QQ}}^+\left(g_{v}^{\Omega_\mathcal{QQ}}\gamma^\mu+g_{t}^{\Omega_\mathcal{QQ}}
\frac{\sigma^{\mu\nu}\partial_\nu}{2m_B}\right)\phi_\mu
\Omega_{\mathcal{QQ}}^{+}.\label{eq:LBBV}
\end{eqnarray}

According to the OZI rule, $\Xi_{\mathcal{QQ}}/\Omega_{\mathcal{QQ}}$
only couples to $(\rho, \omega)/\phi$. Furthermore, given the large breaking
of $SU(3)$ symmetry, we take different values for the couplings of $\Xi_{\mathcal{QQ}}$ and
$\Omega_{\mathcal{QQ}}$.

 The Lagrangian of the vector-meson coupling to the photon, needed
to calculate the contributions of vector mesons to the form
factors, is given by~\cite{Borasoy:1995ds}
\begin{eqnarray}
\mathcal{L}_\gamma=-\frac{1}{2\sqrt{2}}\frac{F_V}{M_V}\langle
V_{\mu\nu} f^{+\mu\nu}\rangle.\label{eq:Lgamma}
\end{eqnarray}
Here, $V_{\mu\nu}=\partial_\mu V_\nu-\partial_\nu V_\mu$ with
$V_\mu$ the 3$\times$3 matrix
\begin{eqnarray}
V_\mu=\left(
        \begin{array}{ccc}
          \frac{1}{\sqrt{2}}\rho^0+\frac{1}{\sqrt{2}}\omega & \rho^+ & K^{*+} \\
          \rho^- & -\frac{1}{\sqrt{2}}\rho^0+\frac{1}{\sqrt{2}}\omega & K^{*0} \\
          K^{*-} & \bar{K}^{*0} & \phi \\
        \end{array}
      \right).
\end{eqnarray}
In Eq. \eqref{eq:Lgamma}, $M_V$ is the mass of the
vector meson. $F_V$  can be obtained by calculating the decay width
$V\to e^+e^-$
\begin{eqnarray}
\Gamma_{V\to e^+e^-}=C_V^2\frac{4\pi \alpha^2F_V^2}{3M_V}
\end{eqnarray}
with $\alpha=\frac{1}{137}$, and
$C_V=1,\frac{1}{3},-\frac{\sqrt{2}}{3}$ for $\rho$, $\omega$ and
$\phi$, respectively.

\section{Form factors of the doubly-heavy baryons}

\subsection{Definitions}
Considering the baryon matrix elements of the electromagnetic vector current, the electromagnetic form factors are defined as
\begin{eqnarray}
\langle B(p_f)|J^\mu (0)|B(p_i)\rangle &=& \bar{u}(p_f)
\left[\gamma^\mu F_1^B(q^2)+\frac{i\sigma^{\mu\nu}q_\nu}{2m_B}
F_2^B(q^2) \right]u(p_i),\label{eq:formfactordefinition}
\end{eqnarray}
where $J^\mu(x)=\sum_{q}e_q\bar{q}(x)\gamma^\mu q(x)$ with $q$
running over
%the light quarks $u, d$ and $s$
the quarks, and $B$ denotes the
baryon $\Xi_{cc}^{++}$, $\Xi_{cc}^{+}$, $\Omega_{cc}^{+}$ or
$\Xi_{bb}^{0}$, $\Xi_{bb}^{-}$, $\Omega_{bb}^{-}$. The
physical mass of the baryon $B$ is given by $m_B$, $e_q$ is the charge of the quark
$q$, and $F^B_1(q^2)$ and $F^B_2(q^2)$ are the Dirac and Pauli form
factors. The Dirac spinor of a baryon with
four-momentum $p^\mu$ and mass $m$ is denoted as $u(p)$. The transferred four-momentum $q^\mu=p_f^\mu-p_i^\mu$ obeys $q^2\leq 0$. The electric and magnetic form factors are defined as
\begin{eqnarray}
G_E^B(q^2)&=&F_1^B(q^2)+\frac{q^2}{4 m_B^2}F_2^B(q^2),\\
G_M^B(q^2)&=&F_1^B(q^2)+F_2^B(q^2).
\end{eqnarray}
Then, the magnetic moment is defined as
\begin{eqnarray}
\mu_B=G_M(0)\frac{e}{2m_B},
\end{eqnarray}
while the charge and magnetic radii of the  baryons can be obtained from the slope of the electric and magnetic form factors
\begin{eqnarray}
\langle r^2_{E,M}\rangle_B=\left.\frac{6}{G_{E,M}^B(0)}\frac{dG_{E,M}^B(q^2)}{dq^2}\right|_{q^2=0}.
\end{eqnarray}
For neutral baryons, an exception is made for the electric radius, which reads
\begin{eqnarray}
\langle r^2_{E}\rangle_B=\left.6\frac{d G_{E}^B(q^2)}{dq^2}\right|_{q^2=0}.
\end{eqnarray}

%%%%%%%%%%%%%%%%%%%%%%%%%%%%%%%%%%%%%%%%%%%%%%%%%%%%%%%%%%%%%%%%%%%%%%%

\begin{figure*}
\centering
\includegraphics[width=0.75\linewidth]{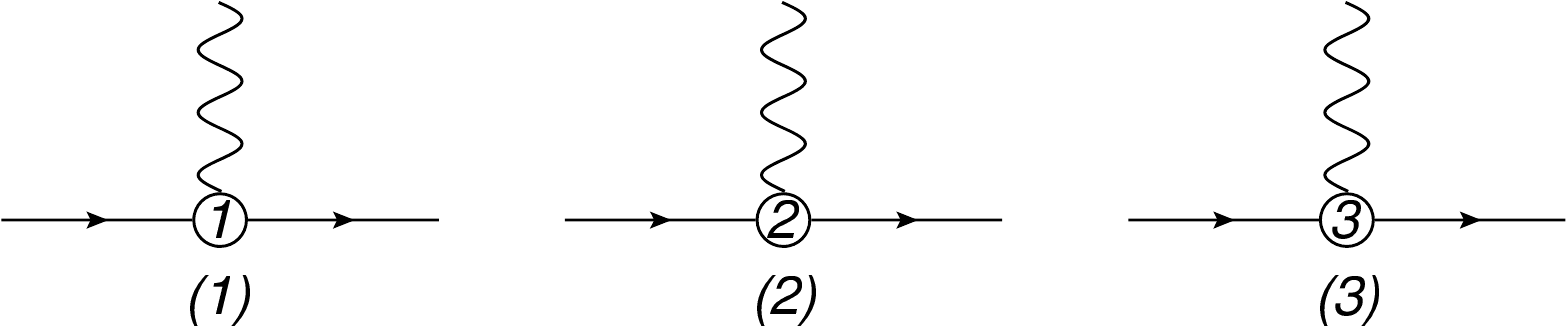}\\
\includegraphics[width=0.55\linewidth]{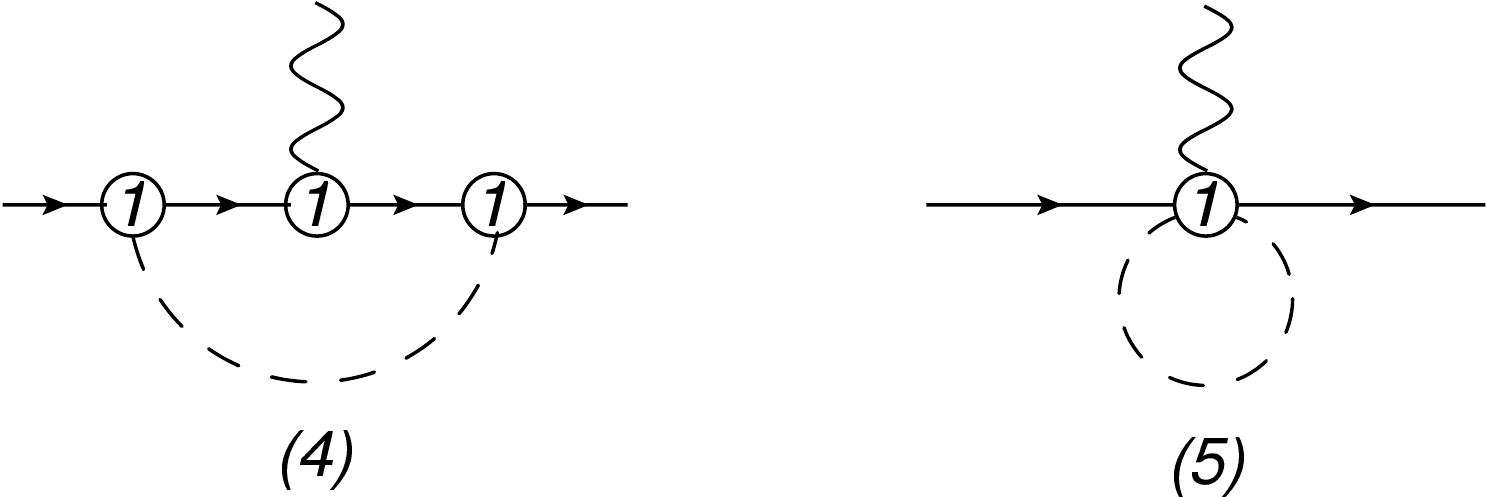}\\
\includegraphics[width=0.55\linewidth]{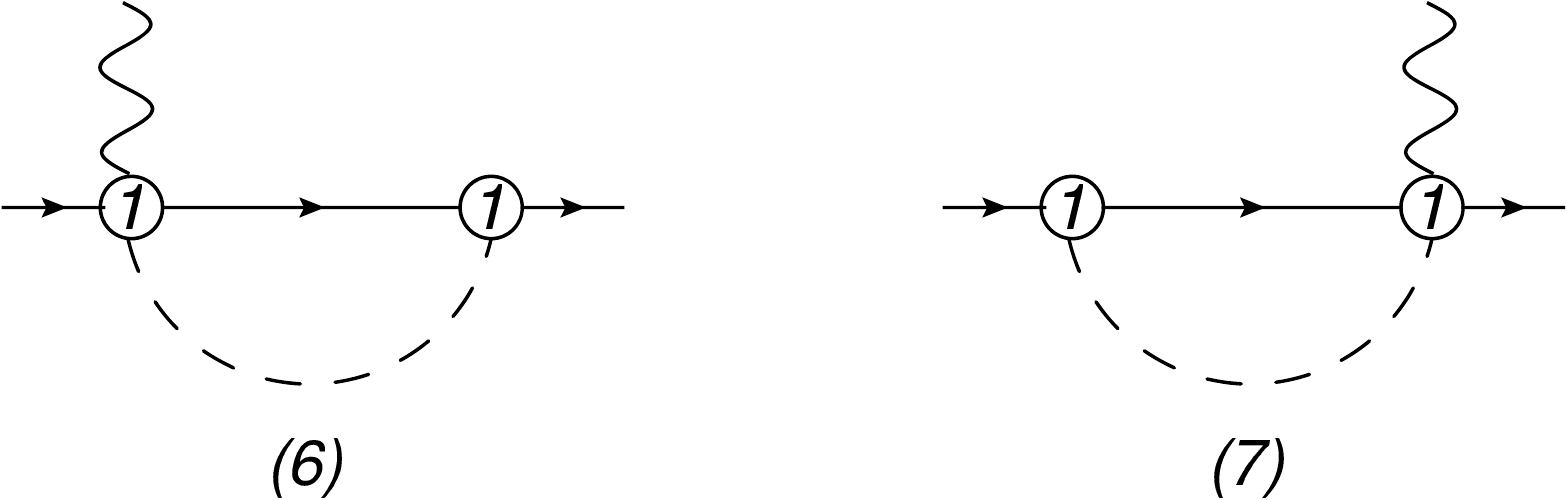}\\
\includegraphics[width=0.55\linewidth]{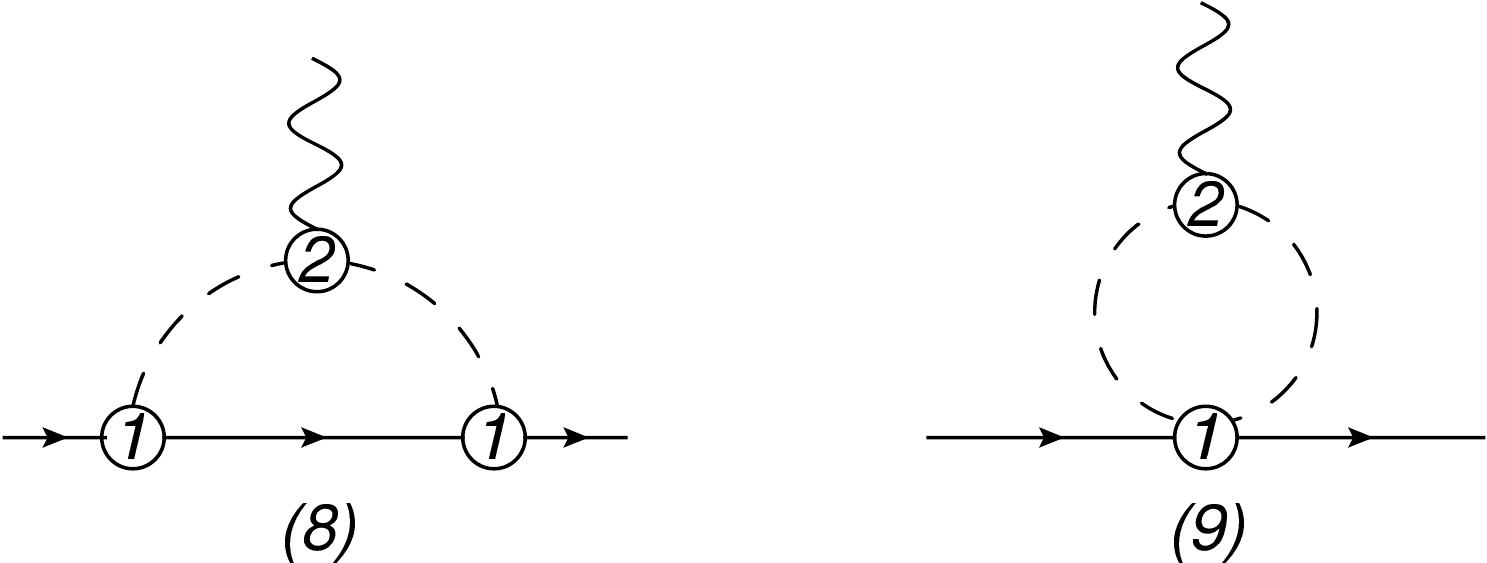}
\caption{Feynman diagrams contributing to the electromagnetic form factors up to order
$O(q^3)$. Wiggled, dashed and solid lines correspond to photon, mesons and baryons respectively. The numbers in the circles show the chiral order of the vertices.
\label{fig:fynmandiagram}}
\end{figure*}

\subsection{Calculation of the form factors}
In Fig.~\ref{fig:fynmandiagram},  we show the diagrams  derived from the Lagrangians of Eqs. \eqref{eq:L1}, \eqref{eq:L2}, \eqref{eq:L3} and \eqref{eq:LM} which contribute to the electromagnetic current matrix element up to order $O(q^3)$ in the chiral expansion. We use the standard ChPT definition for the order of a given diagram~\cite{Scherer:2002tk}. The resulting lengthy expressions of the unrenormalized contributions to the Dirac and Pauli form factors $F_1^B$ and $F_2^B$ are given in the Appendix. We renormalize them following the EOMS scheme. As customary, we perform a modified minimal subtraction ($\widetilde{\textup{MS}}$) \footnote{Namely, we subtract
%Or one can apply the $\overline{\textup{MS}}$ scheme subtracting $1/(4-n)+[\ln(4\pi)+\Gamma^\prime(1)]/2$ instead of
$1/(4-n)+[\ln(4\pi)+\Gamma^\prime(1)+1]/2$, where $n$ is the dimension of dimensional regularization.}. Later, the terms that still break the nominal power counting, which come from the baryonic loops, are also subtracted. In fact, the only subtraction terms required for the Pauli form factors read
\begin{equation}
\Delta F_2^4=C_4\frac{g_A^2m^2}{16\pi^2F^2},\qquad \Delta F_2^8=C_8\frac{g_A^2m^2}{32\pi^2F^2},
\end{equation}
where $C_4$ and $C_8$ are shown in the Appendix.
For the case of the Dirac form factor up to $O(q^3)$, the subtraction vanishes exactly  due to cancellations between diagrams.

To obtain the final expression, one needs to take into account the wave-function renormalization (WFR) given by
\begin{eqnarray}
Z_{\lambda,B}&=&1-C_{\lambda,B}\frac{g_A^2M_\lambda^2}{32\pi^2F_\lambda^2m^2\sqrt{4m^2-M_\lambda^2}}\left\{\sqrt{4m^2-M_\lambda^2}\left[\left(-2+2\ln\frac{m}{\mu}-3\ln\frac{M_\lambda}{\mu}\right)m^2\right.\right.\nonumber\\
&&+2\left.M_\lambda^2\ln\frac{M_\lambda}{m}\right]\left.-2M_\lambda(-3m^2+M_\lambda^2)\left[\arctan\frac{M_\lambda}{\sqrt{4m^2-M_\lambda^2}}\right.\right.\nonumber\\
&&\left.\left.+\arctan\frac{2m^2-M_\lambda^2}{\sqrt{4m^2M_\lambda^2-M_\lambda^4}}\right]\right\}\label{eq:22}.
\end{eqnarray}
Here, the subscript $\lambda$ denotes $\pi^{\pm,0}$, $K^{\pm,0}$, $\bar{K}^0$ or $\eta$, and $M_\lambda$ is the mass of the pseudoscalar meson $\lambda$. The $C_{\lambda,B}$ are shown in Tab. \ref{tab:1}, and a sum over $\lambda$ is inferred.
The WFR constant only multiplies the $\mathcal{O}(q)$ diagrams, since it provides a correction of $\mathcal{O}(q^2)$. Its effect on other diagrams would only start at  $\mathcal{O}(q^4)$, beyond the order of our calculation. Note that a proper inclusion of the WFR is required to ensure that the total baryon charge $F_1(0)=G_E(0)$ is conserved.

\begin{table}[htbp]
\centering
\begin{tabular}{c|ccccccccccc}\toprule[1pt]
$C_{\lambda,B}$&$\pi$&$K$&$\eta$\\\hline
$\Xi_{cc}^{++}/\Xi_{bb}^{0}$&3&2&$\frac{1}{3}$&\\
$\Xi_{cc}^{+}/\Xi_{bb}^{-}$&3&2&$\frac{1}{3}$&\\
$\Omega_{cc}^{+}/\Omega_{bb}^{-}$&0&4&$\frac{4}{3}$&\\\bottomrule[1pt]
\end{tabular}
\caption{Value of the coefficients $C_{\lambda,B}$ in Eq. \eqref{eq:22}.\label{tab:1}}
\end{table}

\subsection{Vector mesons}
 The contributions to the form factors originating from the coupling to the photon through  vector mesons, Fig.~\ref{fig:vector}, are
\begin{eqnarray}
F_1^{VB}&=&-C_{VB}\frac{F_V}{M_V}\frac{g_v^{B}\,q^2}{q^2-M_V^2+i\epsilon},\nonumber\\
F_2^{VB}&=&C_{VB}\frac{F_V}{M_V}\frac{g_t^{B}\,q^2}{q^2-M_V^2+i\epsilon}.\label{eq:FFV}
\end{eqnarray}
In these equations, $B$ denotes the doubly charmed (beauty) baryons $\Xi_{cc}$ ($\Xi_{bb}$) and $\Omega_{cc}$ ($\Omega_{bb}$).
The $C_{VB}$ values are in Tab. \ref{T:CVB}.
Obviously, the mechanism of Fig. \ref{fig:vector} does not contribute to the Pauli and Dirac form factors as $q^2\to 0$. Therefore, the vector mesons do not affect the charge nor the magnetic moment of the baryons.
\begin{table}[ht]
\caption{Values of $C_{VB}$  in Eqs.~\eqref{eq:FFV}.} \label{T:CVB}
\begin{tabular}{c|c|c|c}
&$\Xi_{cc}^{++}/\Xi_{bb}^0$&$\Xi_{cc}^{+}/\Xi_{bb}^-$&$\Omega_{cc}^{+}/\Omega_{bb}^-$\\\hline
$\rho$&$\frac{1}{\sqrt{2}}$&-$\frac{1}{\sqrt{2}}$&0\\
$\omega$&$\frac{1}{3\sqrt{2}}$&$\frac{1}{3\sqrt{2}}$&0\\
$\phi$&0&0&$-\frac{\sqrt{2}}{3}$
\end{tabular}
\end{table}

\section{Results}

For the numerical results presented in this section, we take $m_{\Xi_{cc}}=3.621$
GeV~\cite{Aaij:2017ueg}, $m_{\Omega_{cc}}=3.652$
GeV~\cite{Sun:2016wzh}, $F_\pi=92$ MeV, $F_K=112$ MeV and $F_\eta=110$
MeV.
%$M_\pi=137$ MeV, $M_K=496$ MeV, $M_\eta=548$ MeV and the nucleon mass $m_N =0.938$ GeV.
Furthermore, we set the renormalization scale in the loop diagrams
to 1 GeV, and the chiral limit mass $m$ to the physical baryon mass  $m_B$. The coupling of the pseudoscalar mesons to the doubly charmed baryons is fixed at
$g_A=-0.2$~\cite{Sun:2016wzh}. For $M_\pi$, $M_K$, $M_\eta$, the nucleon mass $m_N$,
the vector meson masses and their widths, we use the averaged  PDG values~\cite{Patrignani:2016xqp}.

In order to estimate the relevant low-energy constants $c_8$, $c_9$, $d_1$ and $d_2$, we use the lattice results from Refs.~\cite{Can:2013tna,Can:2013zpa,Can:2015exa}. There, the magnetic moments and electromagnetic form factors of $\Omega_{cc}^+$ and $\Xi_{cc}^+$ at different values of $Q^2$ were obtained for different lattice configurations, and therefore different meson and baryon masses. Since the scale of ChPT is approximately $\Lambda\sim 1$~GeV, we take into account for the fit the lattice results up to $Q^2<0.4$~GeV$^2$ and $M_\pi^2<0.4$~GeV$^2$, meaning a total of 34 data points.
For the fit, we set the pion and baryon masses in the ChPT calculations to those given by the lattice collaboration. For the kaon and the $\eta$ meson masses, not explicit in Refs.~\cite{Can:2013tna,Can:2013zpa,Can:2015exa},
we use the Gell-Mann, Oakes, and Renner relations~\cite{GellMann:1968rz,Scherer:2002tk}  taking into account that the strange quark mass is fixed to its physical value.

\subsection{Magnetic Moments}

In Table~\ref{T:mucc}, we show our results for the double-charm
baryon magnetic moments $\mu_B$. The vector mesons do not contribute to this observable. The tree diagram contributions are the same for $\Xi_{cc}^{+}$ and $\Omega_{cc}^{+}$.   For the loop terms, we show the analytic expression of the leading-order heavy-baryon expansion, which reproduces
the findings of Ref.~\cite{Li:2017cfz}\footnote{There is a factor 2
discrepancy between our work and Ref.~\cite{Li:2017cfz} in the
vertex definitions involving $g_A$, leading to an overall factor 4
difference in the analytic heavy-baryon results. This simply
translates into different values for $g_A$ when fitting to data.
Indeed, while we use the value $g_A= -0.2$~\cite{Sun:2016wzh}, their
estimate is of $g_A=-0.5$.},  and compare the numerical results in HB with the covariant EOMS scheme. We find appreciable differences between the two schemes, especially for $\Xi_{cc}^{++}$.

\begin{table}[ht]
\caption{Contributions to $\mu_B$ for the double-charm baryons,
split into tree-level and loop terms. The last four columns are in units of $\mu_N$. Best-fit results for the magnetic moments are shown in boldface. Lattice estimations from the quadratic fit of Ref.~\cite{Can:2013tna} are in the last column.} \label{T:mucc}
\begin{tabular}{c|c|c|c|c|c|c}
&Tree&Loops HB&Loop HB $[\mu_N]$&Loop EOMS $[\mu_N]$& $\mu\,[\mu_N]$&Ref.~\cite{Can:2013tna}\\\hline
$\Xi_{cc}^{++}$&$2+\frac{2}{3}c_8+4c_9$   &$-\frac{g_A^2 }{8\pi }\left[\frac{M_\pi m_{\Xi_{cc}}}{F_\pi^2}+\frac{M_Km_{\Omega_{cc}} }{F_K^2}\right]$&-2.09$g_A^2$&$-1.21g_A^2$&---&---\\
$\Xi_{cc}^{+}$&  $1-\frac{1}{3}c_8+4c_9$  &$\frac{g_A^2 m_{\Xi_{cc}}}{8\pi }\frac{M_\pi }{F_\pi^2}$&0.60$g_A^2$&$0.80g_A^2$&{\bf 0.37(2)}&0.425(29)\\
$\Omega_{cc}^{+}$&$1-\frac{1}{3}c_8+4c_9$&$\frac{g_A^2
m_{\Xi_{cc}}}{8\pi }\frac{M_K }{F_K^2}$&1.46$g_A^2$&$1.59g_A^2$&{\bf 0.40(3)}&0.413(24)
\end{tabular}
\end{table}

Up to $\mathcal{O}(q^3)$, the magnetic moments depend only on  $c_8$ and $c_9$  and several known parameters. Furthermore, the available lattice data for magnetic moments correspond to the $\Omega_{cc}^{+}$ and $\Xi_{cc}^{+}$ baryons. For these two particles,  $c_8$ and $c_9$ appear just in the combination $c_{89}=-\frac{1}{3}c_8+4c_9$. Thus, making a fit to the 13 lattice data for magnetic moments, the only free parameter is $c_{89}$, and we can obtain an estimate for this constant: $c_{89}=0.32(2)$.\footnote{In Ref.~\cite{Can:2013zpa}, only the  $\Xi_{cc}$ was studied, and there is no information on the $\Omega_{cc}$ lattice mass needed for the loop contributions. Thus,
for these data we use the values of $m_{\Omega_{cc}}$ given for the same lattice configurations in Ref.~\cite{Can:2013tna}.
Similarly, in Ref.~\cite{Can:2015exa} only the $\Omega_{cc}$ is studied, and  we use
the linear fit from Ref.~\cite{Can:2013tna} for the $\Xi_{cc}$ mass, $m_{\Xi_{cc}}=3.660$~GeV. The sensitivity of the results to these choices is negligible.}
The agreement of our $\mu_{\Xi_{cc}^{+}}$ and $\mu_{\Omega_{cc}^{+}}$ results with the simple extrapolations of the lattice data to the physical point done in Refs.~\cite{Can:2013tna,Can:2013zpa,Can:2015exa} is good considering the uncertainties.

%Nonetheless, there is a somehow larger split, and in the opposite  direction, between the  $\Omega_{cc}^{+}$ and $\Xi_{cc}^{+}$ magnetic moments than in the lattice.  In our model, that splitting is unavoidable and comes from the loops accounting for the mesonic cloud. The HB calculation at the same order would lead to an even larger separation. If the small difference between the  magnetic moments is confirmed by the experiment or further more precise  lattice results, this would indicate the need for a higher order calculation.

 As can be seen,  the loop corrections obtained from the relativistic EOMS renormalization are larger than in the HB approach for $\Xi_{cc}^{+}$ and  $\Omega_{cc}^{+}$. The main reason for this is that most of the loop diagrams in Fig.~\ref{fig:fynmandiagram} enter only at $\mathcal{O}(q^4)$ in HB (only diagram (8)
contributes at $\mathcal{O}(q^3)$), while in the EOMS scheme they are all non-vanishing already at $\mathcal{O}(q^3)$.
%\footnote{Note, though, that at $t=q^2=0$, the loop contributions to $F_1$ are canceled by the WFR  and the only loop diagrams that give contributions to $F_2$, and thus to $\mu_B$, are (4) and (8).}

In these results, the  LEC uncertainties are purely statistical. However,
the chiral error estimates, $\delta\mu$, of the magnetic moments are performed as in Refs.~\cite{Epelbaum:2014efa,Yao:2017fym} and try to account for the systematic error due to the truncation of the chiral series. For our particular case we have
\begin{align}
\delta\mu=\text{max}\left[\mu^{(1)}\left(\frac{M_\pi}{\Lambda}\right)^3,\left\{|\mu^{(k)}-\mu^{(j)}|\left(\frac{M_\pi}{\Lambda}\right)^{3-j}\right\}\right],\quad 1\leq j\leq k\leq 3,\label{eq:unc}
\end{align}
where $\mu^{(i)}$ are the magnetic moments obtained with our best-fit parameters, up to the order $\mathcal{O}(p^i)$.

\begin{table} [ht]
\caption{Contributions to $\mu_B$ for the double-beauty baryons,
split into tree-level and loop terms. For the latter, we show the
analytic expression for the leading-order heavy-baryon (HB)
expansion, and compare the numerical results in HB with the
covariant EOMS scheme, in units of $\mu_N$.} \label{T:mubb}
\begin{tabular}{c|c|c|c|c}
&Tree&Loops HB&Loops HB $[\mu_N]$&Loops EOMS $[\mu_N]$\\\hline
$\Xi_{bb}^{0}$&$\frac{2}{3}\tilde{c}_8-2\tilde{c}_9$   &$-\frac{\tilde{g}_A^2 }{8\pi }\left[\frac{M_\pi m_{\Xi_{bb}} }{F_\pi^2}+\frac{M_K m_{\Omega_{bb}}}{F_K^2}\right]$&-2.11$\tilde{g}_A^2$&$-1.92\tilde{g}_A^2$\\
$\Xi_{bb}^{-}$&  $-1-\frac{1}{3}\tilde{c}_8-2\tilde{c}_9$  &$\frac{\tilde{g}_A^2 m_{\Xi_{bb}}}{8\pi }\frac{M_\pi }{F_\pi^2}$&0.60$\tilde{g}_A^2$&$0.44\tilde{g}_A^2$\\
$\Omega_{bb}^{-}$&$-1-\frac{1}{3}\tilde{c}_8-2\tilde{c}_9$&$\frac{\tilde{g}_A^2
m_{\Xi_{bb}}}{8\pi }\frac{M_K }{F_K^2}$&1.46$\tilde{g}_A^2$&$1.02\tilde{g}_A^2$
\end{tabular}
\end{table}
We show the analogous results for the double-beauty magnetic moments
in Table~\ref{T:mubb}. Again, due to the symmetry, the
tree-level expressions are the same for the two baryons with the same
charge, $\Xi_{bb}^-$ and $\Omega_{bb}^-$. The only difference
between the HB expansions of the charm triplet and the beauty
triplet are the baryon masses. We take $m_{\Xi_{bb}}=10.314$ GeV and $m_{\Omega_{bb}}=10.476$ GeV~\cite{Yoshida:2015tia} for the numerical calculations. Since the double-beauty baryons are substantially
heavier than the double-charm ones, a HB approach is expected to
give a better approximation of the full relativistic result. Indeed, for $\Xi_{bb}^0$ the differences between HB and EOMS results become smaller. However, for the other two baryons the differences are still large, as was the case in the double-charm sector. In the double-beauty sector, all the magnetic moments are systematically of a smaller magnitude when calculated in EOMS than when determined in a HB approach.\footnote{It should be mentioned that the EOMS results are not simply obtained by the substitution of the double-charm baryon masses in the loops by their double-beauty partners. Since the beauty triplet contains a baryon of neutral charge, some of the contributions from diagram (4) in Fig.~\ref{fig:fynmandiagram} to the charm triplet now vanish.}

\subsection{Electric and magnetic radii}
The electric and magnetic radii, $r_{E,M}^B$, measure the derivative with respect to $q^2$ of the
$G_{E,M}^B$ form factors.
%depend on the contributions of the diagrams of Figs.~\ref{fig:vector} and \ref{fig:fynmandiagram} at $t=q^2\neq 0$.
The tree-level results for  $F_1$ and $F_2$ are shown in Tab.~\ref{T:F1cc}.  Since the LECs $c_8$ and $c_9$
appear as the combination $c_{89}$, and $d_1$ and $d_2$ as $d_{12}\equiv-\frac{d_1}{3}+4d_2$ for both the
$\Omega_{cc}^+$ and the $\Xi_{cc}^+$, if we analyze data for only these latter particles, the number of degrees of
freedom from the chiral Lagrangian is reduced to just two.

As it happens for light baryons, we expect the vector mesons to be relevant for these observables. As simplifying conjectures, we assume the OZI rule and ideal
mixing, which implies that the vector-meson contributions to the
$\Xi_{cc}$ form factors come from $\rho$ and $\omega$ and
 $\phi$ is the only vector-meson contributing in the $\Omega_{cc}$ case. Still, this amounts to four unknown parameters: $g_{v}^{\Xi_{cc}}$, $g_{t}^{\Xi_{cc}}$, $g_{v}^{\Omega_{cc}}$ and $g_{t}^{\Omega_{cc}}$. In the magnetic form factors, $g_{v}^{\Xi_{cc}}$ and $g_{t}^{\Xi_{cc}}$ appear as the combination $g_{v}^{\Xi_{cc}}-g_{t}^{\Xi_{cc}}\equiv g_{vt}^{\Xi_{cc}}$. Therefore, in order to separate them, one needs additionally information on the electric form factors. The same is true for $g_{v}^{\Omega_{cc}}$ and $g_{t}^{\Omega_{cc}}$. Since we have lattice results on $G_E^{\Omega^+_{cc}}$, but not on $G_E^{\Xi^+_{cc}}$, we can obtain the values of $g_{v}^{\Omega_{cc}}$ and $g_{t}^{\Omega_{cc}}$ separately, but only the combination $g_{vt}^{\Xi_{cc}}$ can be determined.

\begin{figure}[ht]
\includegraphics[width=.45\textwidth]{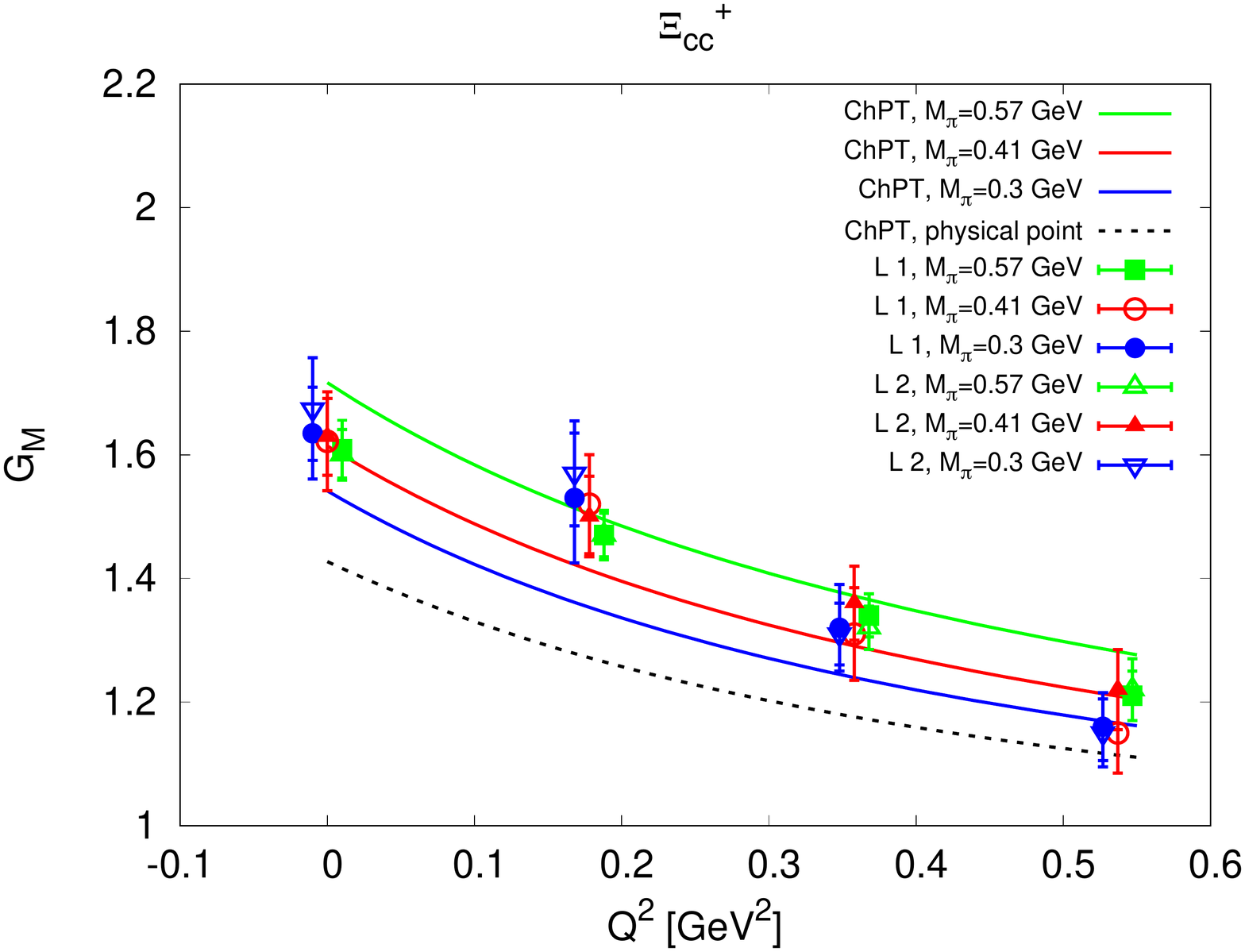}
\includegraphics[width=.45\textwidth]{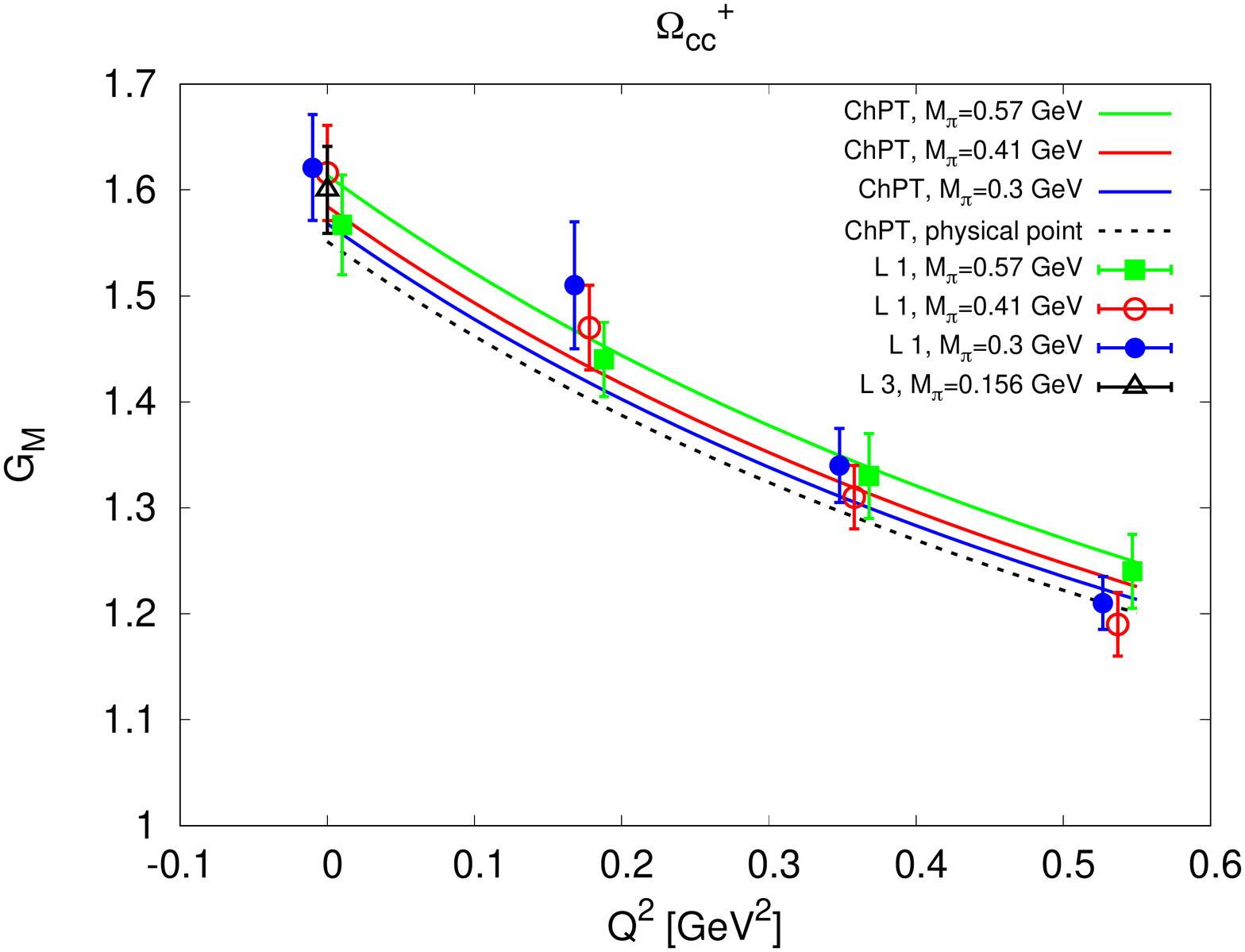}
\includegraphics[width=.45\textwidth]{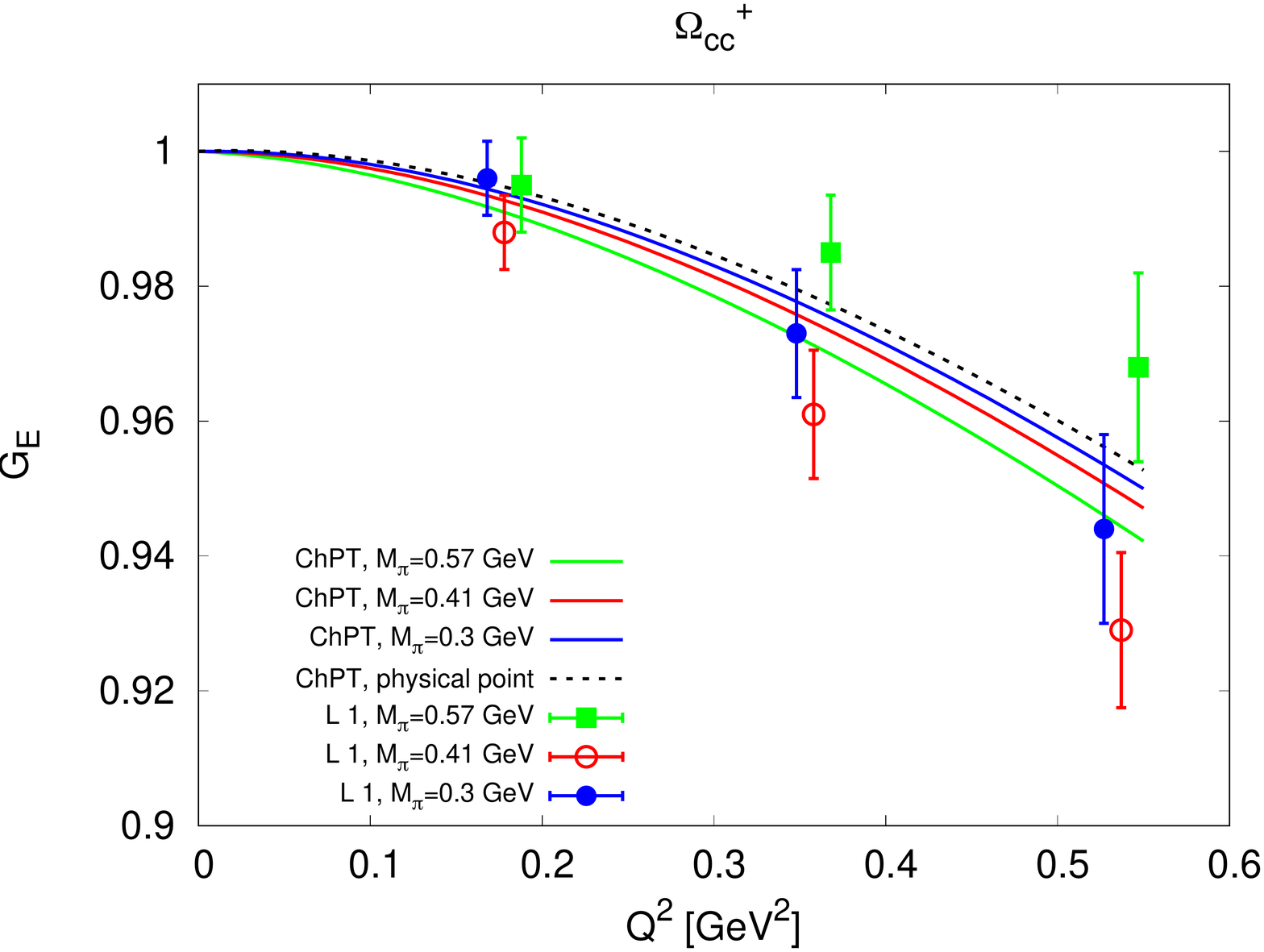}
\caption{Magnetic and electric form factor fit results compared to the lattice data from Refs.~\cite{Can:2013tna}, \cite{Can:2013zpa} and \cite{Can:2015exa}, denoted as L1, L2 and L3, respectively. We show the results for four different pion mass configurations: at the physical point (dotted line), at $M_\pi=0.3$~GeV (blue), $M_\pi=0.41$~GeV (red), $M_\pi=0.57$~GeV (green). The data points for $M_\pi=0.3$~GeV($M_\pi=0.57$~GeV) were shifted to the left(right) for better visibility, but they all correspond to the same $Q^2$ value as those for  $M_\pi=0.41$~GeV.}
\label{Fig:GM_GE}
\end{figure}

Fitting the full set of lattice $G_{E,M}(Q^2)$ results with $Q^2<0.4$~GeV$^2$ and $M_\pi^2<0.4$~GeV$^2$ from \cite{Can:2013zpa,Can:2013tna,Can:2015exa}, we obtain
$c_{89}=0.32(2)$, $d_{12}=(-0.12\pm 0.11)$, $g_{vt}^{\Xi_{cc}}=-10.4(7)$, $g_{v}^{\Omega_{cc}}=(-3.7\pm 3.9)$ and $g_{t}^{\Omega_{cc}}=(-18.9\pm 4.2)$, with a reduced $\chi^2\approx 2.1$.
The value for the parameter $c_{89}$ coincides with the determination obtained using only the magnetic moments. In Fig.~\ref{Fig:GM_GE}, we show our results for the central values of the LECs, compared with the corresponding lattice data. The agreement is fair in the range of $Q^2$ considered. Notice, however, the large uncertainties, not reflected in the figure, in some of the LECs.
\begin{table}
\caption{Tree-level contributions to the double charm $F_1$ and
$F_2$ from the chiral Lagrangian ($\chi$PT) and vector-meson
diagrams (VM).} \label{T:F1cc}
\begin{tabular}{c|c|c|c|c}
&$\chi$PT $F_1$&VM $F_1$&$\chi$PT $F_2$&VM $F_2$\\\hline
$\Xi_{cc}^{++}$&$2-\frac{4 d_1}{3}t-8d_2 t$&$-\displaystyle\sum_{V=\rho,\omega,\phi}C_{VB}\frac{F_V\,t}{M_V}\frac{g_{v}^{\Xi_{cc}}}{t-M_V^2}$&$\frac{2}{3}c_8+4c_9+\frac{4 d_1}{3}t+8d_2 t$ &$\displaystyle\sum_{V=\rho,\omega,\phi}C_{VB}\frac{F_V\,t}{M_V}\frac{g_{t}^{\Xi_{cc}}}{t-M_V^2}$\\
$\Xi_{cc}^{+}$&$1+\frac{2 d_1}{3}t-8d_2 t$&$-\displaystyle\sum_{V=\rho,\omega,\phi}C_{VB}\frac{F_V\,t}{M_V}\frac{g_{v}^{\Xi_{cc}}}{t-M_V^2}$&$-\frac{1}{3}c_8+4c_9-\frac{2 d_1}{3}t+8d_2 t$&$\displaystyle\sum_{V=\rho,\omega,\phi}C_{VB}\frac{F_V\,t}{M_V}\frac{g_{t}^{\Xi_{cc}}}{t-M_V^2}$\\
$\Omega_{cc}^{+}$&$1+\frac{2 d_1 }{3}t-8d_2
t$&$-\displaystyle\sum_{V=\rho,\omega,\phi}C_{VB}\frac{F_V\,t}{M_V}\frac{g_{v}^{\Omega_{cc}}}{t-M_V^2}$&$-\frac{1}{3}c_8+4c_9-\frac{2
d_1 }{3}t+8d_2
t$&$\displaystyle\sum_{V=\rho,\omega,\phi}C_{VB}\frac{F_V\,t}{M_V}\frac{g_{t}^{\Omega_{cc}}}{t-M_V^2}$\end{tabular}
\end{table}
The quality of this $Q^2$ description depends heavily on the inclusion of the vector-meson effects. Indeed, a fit which only includes chiral Lagrangian terms leads to a much higher reduced $\chi^2\approx 9.9$.\footnote{Furthermore, the numerical values of $c_{89}$ and $d_{12}$ would change drastically to $0.09(3)$ and $-0.02(2)$, respectively.} The alternative to the explicit vector-meson consideration would be a calculation at a higher chiral order.

From these electric and magnetic form factors we can extract the corresponding radii. We obtain $<r_E^2>^{\Omega_{cc}^+}=0.00(10)(1)$~fm$^2$,
$<r_M^2>^{\Xi_{cc}^+}=0.18(2)(1)$~fm$^2$ and
$<r_M^2>^{\Omega_{cc}^+}=0.147(92)(1)$~fm$^2$.
 In this case, the error estimates arise mostly due to the large uncertainties of the fitted parameters. The second number in parentheses corresponds to the uncertainty coming from the chiral truncation calculated as in Eq.~\eqref{eq:unc}. These values for the radii support the expectations of the double heavy baryons being substantially smaller than the single heavy ones or the light baryons.

Within their large uncertainties, these radii are compatible with the lattice results from Refs.~\cite{Can:2013zpa,Can:2013tna,Can:2015exa}. It is worth mentioning, though, that the extrapolation to the physical point with a linear or quadratic fit, as done in Refs.~\cite{Can:2013zpa,Can:2013tna,Can:2015exa}, is not expected to give the correct results at the physical point, since the non-trivial behavior due to chiral loops cannot be taken into account.  Specifically, the pion cloud effects, very relevant at low $Q^2$ values, lead to an unavoidable logarithmic dependence on the pion mass and, therefore, to a rapid curvature of the radii when approaching the physical mass, absent in the extrapolations of Refs.~\cite{Can:2013zpa,Can:2013tna,Can:2015exa}.

\begin{table}
\caption{Tree-level contributions to the double-beauty $F_1$ and
$F_2$, from the chiral Lagrangian ($\chi$PT) and vector-meson
diagrams (VM).} \label{T:F1bb}
\begin{tabular}{c|c|c|c|c}
&$\chi$PT $F_1$&VM $F_1$&$\chi$PT $F_2$&VM $F_2$\\\hline
$\Xi_{bb}^{0}$&$-\frac{4 \tilde{d}_1}{3}t+4\tilde{d}_2 t$&$-\displaystyle\sum_{V=\rho,\omega,\phi}C_{VB}\frac{F_V\,t}{M_V}\frac{g_{v}^{\Xi_{bb}}}{t-M_V^2}$&$\frac{2}{3}\tilde{c}_8-2\tilde{c}_9+\frac{4 \tilde{d}_1}{3}t-4\tilde{d}_2 t$ &$\displaystyle\sum_{V=\rho,\omega,\phi}C_{VB}\frac{F_V\,t}{M_V}\frac{g_{t}^{\Xi_{bb}}}{t-M_V^2}$\\
$\Xi_{bb}^{-}$&$-1+\frac{2 \tilde{d}_1}{3}t+4\tilde{d}_2 t$&$-\displaystyle\sum_{V=\rho,\omega,\phi}C_{VB}\frac{F_V\,t}{M_V}\frac{g_{v}^{\Xi_{bb}}}{t-M_V^2}$&$-\frac{1}{3}\tilde{c}_8-2\tilde{c}_9-\frac{2 \tilde{d}_1}{3}t-4\tilde{d}_2 t$&$\displaystyle\sum_{V=\rho,\omega,\phi}C_{VB}\frac{F_V\,t}{M_V}\frac{g_{t}^{\Xi_{bb}}}{t-M_V^2}$\\
$\Omega_{bb}^{-}$&$-1+\frac{2 \tilde{d}_1 }{3}t+4\tilde{d}_2
t$&$-\displaystyle\sum_{V=\rho,\omega,\phi}C_{VB}\frac{F_V\,t}{M_V}\frac{g_{v}^{\Omega_{bb}}}{t-M_V^2}$&$-\frac{1}{3}\tilde{c}_8-2\tilde{c}_9-\frac{2
\tilde{d}_1 }{3}t-4\tilde{d}_2
t$&$\displaystyle\sum_{V=\rho,\omega,\phi}C_{VB}\frac{F_V\,t}{M_V}\frac{g_{t}^{\Omega_{bb}}}{t-M_V^2}$\end{tabular}
\end{table}

For completeness, we also show the analytical expressions of the tree-level contributions to the form factors for the double beauty baryons in Tab.~\ref{T:F1bb}.

\section{summary}

In this work, we have studied the electromagnetic form factors of the doubly heavy baryons within the framework of covariant ChPT up to the chiral order $\mathcal{O}(q^3)$. We have applied the covariant extended on-mass-shell renormalization scheme to generate a systematic power counting. The vector meson contributions have also been included to describe the behavior of form factors at momentum transfer different from zero.

From the Lagrangian constructed in Eqs.~\eqref{eq:L1}, \eqref{eq:L2}, \eqref{eq:L3} and \eqref{eq:LBBV}, we have obtained the Dirac and Pauli form factors, from which we have extracted magnetic moments, charge and magnetic radii. We have also compared our results  with those extracted in HBChPT. We have found that the differences between the loop term contributions to $\mu_B$ in HBChPT and EOMS approaches are around $10\%\sim 70\%$ for the double charm sector and $10\%\sim 40\%$ for the double beauty sector.

In order to obtain first estimates for the LECs, we have fitted our model to the available lattice results. We have found that the vector meson contribution is necessary for a good description. However, the lattice data are not sufficient to fix all the parameters. Instead, since for $\Omega_{cc}^{+}$ and $\Xi_{cc}^{+}$ the LECs $c_8$ and $c_9$ appear as $c_{89}=-\frac{1}{3}c_8+4c_9$, and $d_1$ and $d_2$ as
$d_{12}=-\frac{1}{3}d_1+4d_2$, we fit only these combinations, thus reducing the amount of fitting parameters. We also assume the validity of the OZI rule and ideal mixing to further constrain the vector meson contributions. With these caveats, we obtain $c_{89}=0.32(2)$, $d_{12}=-0.12(11)$, $g_{vt}^{\Xi_{cc}}=-10.4(7)$, $g_{v}^{\Omega_{cc}}=-3.7\pm3.9$ and
$g_{t}^{\Omega_{cc}}=-18.9\pm 4.2$ with a small $\chi^2$. As a consequence, we deduce the values
$\mu_{\Xi_{cc}^+}=0.37(2)\mu_N$ and $\mu_{\Omega_{cc}^+}=0.40(3)\mu_N$ for the magnetic moments.

Once more lattice data are available, a determination of the separate LECs $d_1$, $d_2$, $c_8$ and $c_9$ and of the vector-meson parameters will be possible, together with a better extraction of charge and magnetic radii of the doubly-heavy baryons.

\section*{Acknowledgments}
This research is supported by MINECO (Spain) and the ERDF (European Commission) grants No. FIS2014-51948-C2-2-P, FIS2017-84038-C2-2-P, and SEV-2014-0398, and by the Generalitat Valenciana under Contract PROMETEOII/2014/0068. It is also supported by the Deutsche Forschungsgemeinschaft DFG and by the National Science Foundation for Young Scientists of China under Grants NO. 11705069 and the Fundamental Research Funds for the Central Universities.

\section*{Appendix: Analytical results and loop integrals}

The unrenormalized Dirac form factors corresponding to the diagrams
in Fig. \ref{fig:fynmandiagram} read
\begin{eqnarray}
\label{eq:Ap1}
F^4_1&=&C_4\frac{g_A^2}{4F^2}\Bigg\{(p_i\cdot p_f +m^2)I_{BB}(q^2)+I_B+M^2\Bigg[4I^{00}_{\lambda BB}(q^2)+P^2I^{PP}_{\lambda BB}(q^2)\nonumber\\
&&+q^2I^{qq}_{\lambda BB}(q^2)+\frac{1}{32\pi^2}\Bigg]+2(m+m_B)^2I^{00}_{\lambda BB}(q^2)-(m+m_B)^2[4I^{00}_{\lambda BB}(q^2)\nonumber\\
&&+P^2I^{PP}_{\lambda BB}(q^2)+q^2I^{qq}_{\lambda BB}(q^2)+\frac{1}{32\pi^2}]-P^2[I_{BB}(q^2)+2M^2I^P_{\lambda BB}(q^2)]\nonumber\\
&&+4[p_i\cdot p_f I^{00}_{\lambda BB}(q^2)+p_i\cdot Pp_f\cdot PI^{PP}_{\lambda BB}(q^2)+p_i\cdot qp_f\cdot qI^{qq}_{\lambda BB}(q^2)]\nonumber\\
&&+4m_B(m+m_B)[I^{00}_{\lambda BB}(q^2)+q^2I^{qq}_{\lambda BB}(q^2)]+8m_B^2(m+m_B)^2I^{PP}_{\lambda BB}(q^2)\nonumber\\
&&+2(m+m_B)m_B[I_{BB}(q^2)+2M^2I^P_{\lambda BB}(q^2)]-8(m+m_B)m_B
\nonumber\\&&\times[I^{00}_{\lambda BB}(q^2)+p_i\cdot PI^{PP}_{\lambda BB}(q^2)-p_i\cdot qI^{qq}_{\lambda BB}(q^2)]\Bigg\},\nonumber\\
F_1^5&=&C_5\frac{1}{8F^2}I_\lambda,\nonumber\\
F^6_1&=&F^7_1=C_{67}\frac{g_A^2}{4F^2}\left\{-I_B-M^2I_{\lambda B}(m_B^2)-(m-m_B)m_BI^p_{\lambda B}(m_B^2)\right\},\nonumber\\
F^8_1&=&C_8\frac{g_A^2}{8F^2}\left\{-4m_B(m+m_B)I^{00}_{\lambda\lambda B}(q^2)-2q^2I^{00}_{\lambda\lambda}(q^2)+2(m_B^2-m^2)I^{00}_{\lambda\lambda B}(q^2)\right\}-F^8_2,\nonumber\\
F^9_1&=&C_9\frac{1}{4F^2}q^2I^{00}_{\lambda\lambda}(q^2),\\
F^4_2&=&-C_4\frac{g_A^2}{4F^2}\left\{4(m+m_B)^2m_B^2I^{PP}_{\lambda BB}(q^2)+(m+m_B)m_B[I_{BB}(q^2)+2M^2I^P_{\lambda BB}(q^2)]\right.\nonumber\\
&&-4(m+m_B)m_B\left.[I^{00}_{\lambda BB}(q^2)+p_i\cdot PI^{PP}_{\lambda BB}(q^2)-p_i\cdot q I^{qq}_{\lambda BB}(q^2)]\right\},\nonumber\\
F^5_2&=&F^6_2=F^7_2=0,\nonumber\\
F^8_2&=&C_8\frac{g_A^2}{8F^2}\left\{8m_B^2(m+m_B)[2m_BI^{PP}_{\lambda \lambda B}(q^2)
%\right.\nonumber\\
+m_BI^P_{\lambda\lambda B}(q^2)]-8m_B^3(m+m_B)I^{P}_{\lambda \lambda B}(q^2)\right.\nonumber\\
&&-2m_B(m+m_B)[4I^{00}_{\lambda\lambda B}(q^2)+(4m_B^2-q^2)I^{PP}_{\lambda\lambda B}(q^2)+q^2I^{qq}_{\lambda\lambda B}(q^2)+\frac{1}{32\pi^2}]\nonumber\\
&&-8(m+m_B)m_B[I^{00}_{\lambda\lambda B}(q^2)+(2m_B^2-q^2/2)I^{PP}_{\lambda\lambda B}(q^2)+\frac{q^2}{2}I^{qq}_{\lambda\lambda B}(q^2)+m_B^2I^P_{\lambda\lambda B}(q^2)]\nonumber\\
&&+4(m+m_B)m_B[(2m_B^2-\frac{q^2}{2})I^P_{\lambda\lambda B}(q^2)+\frac{q^2}{4}I_{\lambda\lambda B}(q^2)]\nonumber\\
&&+4(m+m_B)m_B[I_{\lambda B}(m_B^2)+M^2(I^P_{\lambda\lambda B}(q^2)+\frac{1}{2}I_{\lambda\lambda B}(q^2))]\nonumber\\
&&-2m_B[m_BI^p_{\lambda B}(m_B^2)+2M^2m_BI^P_{\lambda\lambda B}(q^2)]
-4m_B(m_B^2-m^2)[2m_BI^{PP}_{\lambda\lambda B}(q^2)+m_BI^{P}_{\lambda\lambda B}(q^2)]\nonumber\\
&&\left.+4m_B[m_BI^{00}_{\lambda\lambda B}(q^2)+2m_B(2m_B^2-q^2/2)I^{PP}_{\lambda\lambda B}(q^2)+\frac{m_Bq^2}{2}I^P_{\lambda\lambda B}(q^2)]\right\},\nonumber\\
F^9_2&=&0.\nonumber
\end{eqnarray}
In these equations,
the lower index of the $F$'s is 1(2) for the Dirac(Pauli) form factor. The upper index corresponds to the Fig.~\ref{fig:fynmandiagram} label.
The values of the constants $C_4$, $C_5$, $C_{67}$, $C_8$ and $C_9$ are listed in
Tab. \ref{T:CValues} for the different baryons and mesons. A sum over the mesons ($\lambda$-subindex) is understood.
Also, $m$ is the doubly heavy baryon mass in the chiral limit, $m_B$ is the corresponding physical mass, $q_\mu$ is the four momentum of the photon, $p_{i\mu}$ and $p_{f\mu}$ are the four momenta of the initial and final doubly heavy baryons, and $P_\mu$ is defined as $P_\mu=p_{i\mu}+p_{f\mu}$.

\begin{table}[ht]
\caption{Values of the parameters $C_4$, $C_5$, $C_{67}$, $C_8$, $C_9$ of Eq.~\eqref{eq:Ap1}.}
\label{T:CValues}
\begin{tabular}{c|ccc||ccc||ccc||ccc||cccccccccccccccccccccc}
               &     &$C_4$&      &     &$C_5$&&&$C_{67}$&&&$C_8$&&&$C_9$&&\\\hline
               &$\pi$&$K$  &$\eta$&$\pi$&$K$&$\eta$&$\pi$&$K$&$\eta$&$\pi$&$K$&$\eta$&$\pi$&$K$&$\eta$\\
$\Xi_{cc}^{++}$&4&2&$\frac{2}{3}$&-4&-4&0&-2&-2&0&4&4&0&4&4&0\\
$\Xi_{cc}^{+}$&5&2&$\frac{1}{3}$&4&0&0&2&0&0&-4&0&0&-4&0&0\\
$\Omega_{cc}^{+}$&0&6&$\frac{4}{3}$&0&4&0&0&2&0&0&-4&0&0&-4&0\\
$\Xi_{bb}^{0}$&-2&-2&0&-4&-4&0&-2&-2&0&4&4&0&4&4&0\\
$\Xi_{bb}^{-}$&-1&-2&$-\frac{1}{3}$&4&0&0&2&0&0&-4&0&0&-4&0&0\\
$\Omega_{bb}^{-}$&0&-2&$-\frac{4}{3}$&0&4&0&0&2&0&0&-4&0&0&-4&0\\
\end{tabular}
\end{table}

The loop integrals of Eq.~\eqref{eq:Ap1} are given below.

\begin{eqnarray}
I_B &=& i\mu^{4-n}\int \frac{d^nk}{(2\pi)^n}\frac{1}{k^2-m^2+i\epsilon}=2m^2\left\{R+\frac{1}{32\pi^2}\ln\frac{m^2}{\mu^2}\right\},\nonumber\\
I_\lambda &=& i\mu^{4-n}\int \frac{d^nk}{(2\pi)^n}\frac{1}{k^2-M_\lambda^2+i\epsilon}=2M_\lambda^2R+\frac{M_\lambda^2}{(4\pi)^2}\ln\frac{M_\lambda^2}{\mu^2},\nonumber\\
I_{\lambda B}(p^2)&=&i\mu^{4-n}\int \frac{d^nk}{(2\pi)^n}\frac{1}{(p-k)^2-m^2+i\epsilon}\frac{1}{k^2-M_\lambda^2+i\epsilon}\nonumber\\
&=&2\left\{R+\frac{1}{32\pi^2}\ln\frac{m^2}{\mu^2}\right\}+\frac{1}{(4\pi)^2}\left(-1+\frac{p^2-m^2+M_\lambda^2}{2p^2}\ln \frac{M_\lambda^2}{m^2}+f_0\right),\nonumber\\
I_{\lambda\lambda}(q^2)&=&i\mu^{4-n}\int \frac{d^nk}{(2\pi)^n}\frac{1}{(q+k)^2-M_\lambda^2+i\epsilon}\frac{1}{k^2-M_\lambda^2+i\epsilon}=2R+\frac{1}{16\pi^2}\left[1+2\ln\frac{M}{\mu}+f_0^\prime\left(\frac{q^2}{M^2}\right) \right],\nonumber\\
I_{BB}(q^2)&=&i\mu^{4-n}\int \frac{d^nk}{(2\pi)^n}\frac{1}{(q+k)^2-m^2+i\epsilon}\frac{1}{k^2-m^2+i\epsilon}=2\left\{R+\frac{1}{32\pi^2}\ln\frac{m^2}{\mu^2}\right\}+\frac{1}{16\pi^2}\left[1+f_0^\prime\left(\frac{q^2}{m^2}\right) \right],\nonumber\\
I_{\lambda\lambda}^\mu(q) &=&i\mu^{4-n}\int \frac{d^nk}{(2\pi)^n}\frac{k^\mu}{[(q+k)^2-M_\lambda^2+i\epsilon](k^2-M_\lambda^2+i\epsilon)}=q^\mu I^q_{\lambda\lambda}(q^2),\nonumber\\
I_{\lambda\lambda}^{\mu\nu}(q) &=&i\mu^{4-n}\int \frac{d^nk}{(2\pi)^n}\frac{k^\mu k^\nu}{[(q+k)^2-M_\lambda^2+i\epsilon](k^2-M_\lambda^2+i\epsilon)}=g^{\mu\nu}q^2I^{00}_{\lambda\lambda}(q^2)
+q^\mu q^\nu I^{qq}_{\lambda\lambda}(q^2),\nonumber\\
I_{\lambda B}^\mu(p)&=&i\mu^{4-n}\int \frac{d^nk}{(2\pi)^n}\frac{k^\mu}{(p-k)^2-m^2+i\epsilon}\frac{1}{k^2-M_\lambda^2+i\epsilon}=p^\mu I^p_{\lambda B}(p^2),
\end{eqnarray}
where

\begin{eqnarray}
R&=&\frac{1}{(4\pi)^2}\left[-\frac{1}{\epsilon}-\frac{1}{2}\left(\ln \frac{4\pi}{m^2}+1-\gamma\right)\right],\nonumber\\
f_0&=&\left\{
\begin{array}{c}
\frac{\sqrt{\Theta^2-\Delta^2}}{p^2}\arccos \frac{-\Delta}{\Theta},\,-1<\frac{\Delta}{\Theta}<1,\\
\frac{\sqrt{\Delta^2-\Theta^2}}{2p^2}\ln\frac{\Delta+\sqrt{\Delta^2-\Theta^2}}{\Delta-\sqrt{\Delta^2-\Theta^2}}, \, \frac{\Delta}{\Theta}<-1,\\
\frac{\sqrt{\Delta^2-\Theta^2}}{2p^2}\ln\frac{\Delta+\sqrt{\Delta^2-\Theta^2}}{\Delta-\sqrt{\Delta^2-\Theta^2}}-i\pi\frac{\sqrt{\Delta^2-\Theta^2}}{p^2}, \, \frac{\Delta}{\Theta}>1 ,\\
0,\, \frac{\Delta}{\Theta}=\pm 1,
\end{array} \right.\nonumber\\
%%%
%%%
f_0^\prime(x) &=&\left\{
\begin{array}{c}
-2-\sigma \ln\left(\frac{\sigma-1}{\sigma+1}\right), x<0,\\
-2+2\sqrt{\frac{4}{x}-1}\ \text{arccot}\left(\sqrt{\frac{4}{x}-1}\right), 0\leq x \leq 4,\\
-2-\sigma \ln\left(\frac{\sigma-1}{\sigma+1}\right)-i\pi \sigma,
x>4,
\end{array} \right.\nonumber\\
I_{\lambda\lambda}^q(q^2)&=&-\frac{1}{2}I_{\lambda\lambda}(q^2),\nonumber\\
I_{\lambda\lambda}^{qq}(q^2)&=&\frac{1}{t}\left[\frac{1}{3}I_\lambda +\frac{1}{3}(t-M^2_\lambda)I_{\lambda\lambda}(q^2)+\frac{1}{144\pi^2}\left(3M_\lambda^2-\frac{t}{2}\right)\right],\nonumber\\
I_{\lambda\lambda}^{00}(q^2)&=&\frac{1}{t}\left[\frac{1}{6}I_\lambda +\frac{1}{12}(4M_\lambda^2-t)I_{\lambda\lambda}(q^2)-\frac{1}{8\pi^2}\left(\frac{M_\lambda^2}{6}-\frac{t}{36}\right)\right],\nonumber\\
I^p_{\lambda B}(p^2)&=&\frac{1}{2p^2}[I_N-I_\lambda+(p^2-m^2+M_\lambda^2)I_{\lambda B}(p^2)].
\end{eqnarray}
Here $\Delta=p^2-m^2-M_\lambda^2$, $\Theta=2mM_\lambda$,
$\sigma=\sqrt{1-\frac{4}{x}}$ with $x\notin [0,4]$, and $p_\mu$ is an arbitrary four momentum ($p^2\neq 0$), which, in our case, corresponds to the four momentum of the initial or final doubly heavy baryon.

Furthermore, in the following loop integrals corresponding to three internal lines, the on-shell condition (i.e. $p_i^2=p_f^2=m_B^2$) is assumed.
\begin{eqnarray}
I_{\lambda\lambda B}(q^2)&=&i\mu^{4-n}\int
\frac{d^nk}{(2\pi)^n}\frac{1}{(p_i-k)^2-m^2+i\epsilon}\frac{1}{k^2-M_\lambda^2+i\epsilon}\frac{1}{(k+q)^2-M_\lambda^2+i\epsilon},\nonumber\\
I_{\lambda\lambda B}^\mu(q,P)&=&i\mu^{4-n}\int \frac{d^nk}{(2\pi)^n}\frac{k^\mu}{(p_i-k)^2-m^2+i\epsilon}\frac{1}{k^2-M_\lambda^2+i\epsilon}\frac{1}{(k+q)^2-M_\lambda^2+i\epsilon},\nonumber\\
I_{\lambda\lambda B}^{\mu\nu}(q,P)&=&i\mu^{4-n}\int \frac{d^nk}{(2\pi)^n}\frac{k^\mu k^\nu}{(p_i-k)^2-m^2+i\epsilon}\frac{1}{k^2-M_\lambda^2+i\epsilon}\frac{1}{(k+q)^2-M_\lambda^2+i\epsilon},\nonumber\\
I_{\lambda B B}(q^2)&=&i\mu^{4-n}\int \frac{d^nk}{(2\pi)^n}\frac{1}{(k-p_i)^2-m^2+i\epsilon}\frac{1}{k^2-M_\lambda^2+i\epsilon}\frac{1}{(k-p_f)^2-m^2+i\epsilon},\nonumber\\
I_{\lambda B B}^\mu(q,P)&=&i\mu^{4-n}\int \frac{d^nk}{(2\pi)^n}\frac{k^\mu}{(k-p_i)^2-m^2+i\epsilon}\frac{1}{k^2-M_\lambda^2+i\epsilon}\frac{1}{(k-p_f)^2-m^2+i\epsilon},\nonumber\\
I_{\lambda B B}^{\mu\nu}(q,P)&=&i\mu^{4-n}\int \frac{d^nk}{(2\pi)^n}\frac{k^\mu k^\nu}{(k-p_i)^2-m^2+i\epsilon}\frac{1}{k^2-M_\lambda^2+i\epsilon}\frac{1}{(k-p_f)^2-m^2+i\epsilon}.
\end{eqnarray}

The tensor integrals defined above can be reduced to the scalar ones as follows
\begin{eqnarray}
I_{\lambda B B}^{\mu}(q,P)&=&P^\mu I^P_{\lambda BB}(q^2),\nonumber\\
I_{\lambda B B}^{\mu\nu}(q,P)&=&g^{\mu\nu} I^{00}_{\lambda B B}(q^2)+P^\mu P^\nu I^{PP}_{\lambda B B}(q^2)+q^\mu q^\nu I^{qq}_{\lambda B B}(q^2),\nonumber\\
I_{\lambda \lambda B}^\mu(q,P)&=&P^\mu I^{P}_{\lambda \lambda B}(q^2)-\frac{1}{2}q^\mu I_{\lambda \lambda B}(q^2),\nonumber\\
I_{\lambda \lambda B}^{\mu\nu}(q,P)&=&g^{\mu\nu} I^{00}_{\lambda \lambda B}(q^2)+P^\mu P^\nu I^{PP}_{\lambda \lambda B}(q^2)+q^\mu q^\nu I^{qq}_{\lambda \lambda B}(q^2)-\frac{q^\mu P^\nu +P^\mu q^\nu}{2}I^P_{\lambda \lambda B}(q^2),
\end{eqnarray}
where
\begin{eqnarray}
I^P_{\lambda\lambda B}(q^2)&=&\frac{1}{2(4m_B^2-q^2)}[(2M_\lambda^2-q^2)I_{\lambda\lambda B}(q^2)+2I_{\lambda B}(m_B^2)-2I_{\lambda\lambda}(q^2)],\nonumber\\
I^{00}_{\lambda\lambda B}(q^2)&=&\frac{1}{8}\left[2I_{\lambda B}(m_B^2)+(4M_\lambda^2-q^2)I_{\lambda\lambda B}(q^2)-2(2M_\lambda^2-q^2)I^P_{\lambda\lambda B}(q^2)-\frac{1}{8\pi^2} \right],\nonumber\\
I^{PP}_{\lambda\lambda B}(q^2)&=&\frac{1}{8(4m_B^2-q^2)}\left[-2I_{\lambda
B}(m_B^2)+4I^p_{\lambda B}(m_B^2)-(4M_\lambda^2-q^2)I_{\lambda\lambda B}(q^2)
+6(2M_\lambda^2-q^2)I^P_{\lambda\lambda B}(q^2)+\frac{1}{8\pi^2}\right],\nonumber\\
I^{qq}_{\lambda\lambda B}(q^2)&=&\frac{1}{8q^2}\left[2I_{\lambda B}(m_B^2)-4I^p_{\lambda B}(m_B^2)-(4M_\lambda^2-3q^2)I_{\lambda\lambda B}(q^2)+2(2M_\lambda^2-q^2)I^P_{\lambda\lambda B}(q^2)+\frac{1}{8\pi^2}\right],\nonumber\\
I^P_{\lambda BB}(q^2)&=&\frac{1}{4m^2-q^2}[M_\lambda^2I_{\lambda BB}(q^2)-I_{\lambda B}(m_B^2)+I_{BB}(q^2)],\nonumber\\
I^{00}_{\lambda BB}(q^2)&=&\frac{1}{2}\left[M_\lambda^2(I_{\lambda BB}(q^2)-I_{\lambda BB}^P(q^2))+\frac{1}{2} I_{BB}(q^2)-\frac{1}{32\pi^2}\right],\nonumber\\
I^{PP}_{\lambda BB}(q^2)&=&\frac{1}{2(4m_B^2-q^2)}\left\{[3I^P_{\lambda BB}(q^2)-I_{\lambda BB}(q^2)]M_\lambda^2-I^p_{\lambda B}(m_B^2)+\frac{1}{2}I_{BB}(q^2)+\frac{1}{32\pi^2} \right\},\nonumber\\
I^{qq}_{\lambda BB}(q^2)&=&\frac{1}{2q^2}\left\{[I^P_{\lambda BB}(q^2)-I_{\lambda BB}(q^2)]M_\lambda^2+I^p_{\lambda B}(m_B^2)-\frac{1}{2}I_{BB}(q^2)+\frac{1}{32\pi^2} \right\}.
\end{eqnarray}

\bibliographystyle{plain}

\end{document}